\documentclass[a4paper]{jpconf}
\usepackage{graphicx}
\usepackage{amsmath}
\usepackage{fancyhdr}
\usepackage{fancyhdr}
\begin{document}

\title{Primordial Black Holes And Gravitational Waves Based On No-Scale Supergravity}

\author{Ioanna D. Stamou}

\address{National and Kapodistrian University of Athens, Department of Physics,  Section of Nuclear {\rm \&} Particle Physics,  GR--15784 Athens, Greece}

\ead{joanstam@phys.uoa.gr}

\begin{abstract}
In this paper we present a class of models in order to explain the production of Primordial Black Holes (PBHs) and Gravitational Waves (GWs) in the Universe. These models are based on no-scale theory.  By breaking the  SU(2,1)/SU(2)$\times$U(1) symmetry we fix one of the two chiral fields and we derive effective scalar potentials which are capable of generating PBHs and GWs. As it is known in the literature there is an important unification of the no-scale models, which leads to the Starobinsky effective scalar potential based on the coset  SU(2,1)/SU(2)$\times$U(1). We use this unification in order to  additionally explain the generation of PBHs and GWs. Concretely, we modify well-known superpotentials, which reduce to Starobinsky-like effective scalar potentials. Thus, we derive scalar potentials which, on the one hand, explain the production of PBHs and GWs and, on the other hand, they conserve the transformation laws, which yield from the parametrization of the coset SU(2,1)/SU(2)$\times$U(1) as well as  the unification between the models which are yielded  this coset. We numerically evaluate the scalar power spectra with the effective scalar potential based on this theory. Furthermore, we evaluate the fractional abundances of PBHs by comparing two methods the Press–Schechter approach and the peak theory, while focusing on explaining the dark matter in the Universe. By using the resulting scalar power spectrum we evaluate  the amount of GWs.  All models are in complete consistence with Planck constraints.
\end{abstract}
\section{Introduction}

The detection of Gravitational Waves (GWs)  by a binary black hole merge, reported by LIGO/VIRGO collaboration \cite{Abbott:2016blz,Abbott:2016nmj,Abbott:2017vtc}, opens a new window in physics of Primordial Black Holes (PBHs). As a result there are numerous  recent studies which show that the origin of PBHs  can explain a fraction of Dark Matter (DM) in the Universe. Moreover as the detection of GWs is regarded as a milestone in  Cosmology, there are signals of GWs that are expected to be detected by future space-based interferometer, such as LISA, BBO and DECIGO.  A  well-known way,  in order to  explain both the generation of PBHs and GWs, is the models embedded in inflation. Specifically, an enhancement in scalar power spectrum can be interpreted as a significant amount of both PBHs and GWs, which are formed in the radiation dominated era with the origin to the previous epoch of inflation.

There are numerous  theoretical models, proposed in the literature for explaining the generation of PBHs and GWs. All these models have to deal with the observable constraints released by Planck collaboration\cite{Akrami:2018odb} for the prediction of spectral index $n_s$ and  tensor-to-scalar ratio $r$.  A model which is in complete accordance with this observable bounds is the Starobinsky potential. A theory which can lead to Starobinsky potential is the no-scale supergravity \cite{Ellis:2013xoa,Ellis:2018zya,Ellis:2013nxa}. For this reason we choose to work in the framework of this theory.

Unfortunately, the Starobinsky potential cannot lead to an enhancement of the scalar power spectrum. In the literature many mechanisms have been proposed in order to achieve this amplification. One of the most discussed is an inflection point in the framework of single field inflation. This inflection point can be given as a feature in the effective scalar potential where the first derivative and the second derivative of the potential with respect to the field are almost zero. This mechanism is regarded as a simple way which gives an enhancement in scalar power spectrum. However,
a disadvantage of potentials with an inflection point is that  a lot of fine-tuning of the parameters is required. In our analysis we adopt this mechanism and we analyze the issue of fine-tuning. 


This paper is organized as follows. In the section 2 we  present our theory, which is based on no-scale supergravity, in order to achieve an inflection point in the effective scalar potential. In section 3 we show how our modifications lead to a significant enhancement in the scalar power spectrum. In section 4 we show how this enhancement can be interpreted as the production of PBHs. In section 5 we calculate the energy density of induced GWs. In section 6 we discuss about the issue of the fine-tuning and in section 7 we present some conclusions of our work
\section{Theory}
As the theory of our analysis is based on the SU(2,1)/SU(2)$\times $U(1) symmetry, we briefly refer to some basic aspects about symmetries in the kinetic term in the Lagrangian.
In $N=1$ supergravity the general Lagrangian in effective field theory is given as: 
\begin{equation}
\mathcal{L}=  K^{\bar{j}}_{i} \partial_{\mu} \Phi^i \partial ^{\mu} \bar{\Phi} _{\bar{j}}-V(\Phi)
\end{equation}
where $\Phi_i$ are the chiral fields, $\bar{\Phi_i}$ is its conjugate and subscripts $i,j$ running over the fields. With $K^{\bar{j}}_{i}$ we denote the  K\"ahler metric and $K$ is the K\"ahler potential. We work in Planck units throughout this work, $M_P=1$.
The F-term of scalar potential is given as follows: 
\begin{equation} 
V= e^{K} (D_{\Phi}WK^{\bar{\Phi} \Phi} D_{\bar{\Phi}}\bar{W} - 3|W|^2)
\label{eq2}
\end{equation}
where $W$ is the superpotential and  the K\"ahler covariant derivative is  given:
\noindent
\begin{center}
$D_{\Phi}W=\frac{\partial W}{\partial \Phi}+ \frac{\partial K}{\partial \Phi}W .$
\end{center}
In order to vanish the cosmological constant, or in other words in order to vanish the term $3|W|^2$ in the effective scalar potential, Eq.~(\ref{eq2}), we need to have the following identity:
\begin{equation}
K^{\Phi\bar{\Phi}}K_{\Phi}K_{\bar{\Phi}}=3.
\label{eq4}
\end{equation}
There is an elegant way to obtain this identity by considering the K\"ahler potential in this form \cite{Cremmer:1983bf}:
\begin{equation}
K=-3\ln(\Phi+\bar{\Phi}).
\label{konlyphi}
\end{equation}
This K\"ahler potential represent the minimal no-scale model written in term of one chiral field.
The corresponding kinetic term of the Lagrangian is:
\begin{center}
$\mathcal{L}_{kin}=\frac{3}{(\Phi+\bar{\Phi})}\partial^{\mu}\Phi\partial_{\mu}\bar{\Phi}$.
\end{center}
In order to discuss the symmetries associated with the kinetic term of the Langrangian we consider that the field transforms as $\Phi=(y+1)/(y-1)$. So the K\"ahler potential in the term of the new fiels $y$ is\cite{Ellis:1983ei}:
\begin{equation} 
K=-3 \ln(1-\frac{|y|^2}{3}).
\label{konlyy}
\end{equation}
In this form the corresponding kinetic term of the Lagrangian is given from:
\begin{center}
$\mathcal{L}_{kin}= \frac{3}{(1-|y|^2)^2}\partial^{\mu}y \partial_{\mu} \bar{y}$
\end{center}
which is invariant under the transformation of:
\begin{equation}
y \rightarrow \frac{\alpha y+\beta}{\bar{\beta}y+\bar{\alpha}},
\quad |\alpha|^2-|\beta|^2=1.\end{equation}
This invariance defines the non-compact group SU(1,1). Therefore the K\"ahler potentials written in $T$ basis, from Eq.~(\ref{konlyphi}), and in $y$ basis, from Eq.~(\ref{konlyy}), are two forms of the non-minimal no-scale SU(1,1) model. These equivalent forms of the K\"ahler potential do not lead to Starobinsky-like effective scalar potential. So we need to consider an extension of this group, which is the non-compact 
 SU(2,1)/SU(2)$\times$U(1) coset. Under this coset we consider two chiral fields,  one plays the role of the inflaton and the other is the modulo field.
 
As in the case of SU(1,1), there are two equivalent form of SU(2,1)/SU(2)$\times$U(1) symmetry, which read as:
\begin{equation}
K=-3\ln(1-\frac{|y_1|^2}{3}-\frac{|y_2|^2}{3}) \quad \& \quad K=-3 \ln(T +\bar{T}- \frac{|\varphi|^2}{3}).
\end{equation}
The complex fields $(y_1,y_2)$ are related to $(T,\varphi)$ by the following expressions:
 \begin{center}
 $y_1=\Big(\frac{2\varphi}{1+2T}\Big) ,\quad y_2=\sqrt{3}\Big(\frac{1-2T}{1+2T}\Big)$ 
 \label{k2(2)}
 \end{center}
 \noindent
and the inverse relations by:
 \begin{center}
 $T=\frac{1}{2}\Big(\frac{1 -y_2 / \sqrt{3}}{1 +y_2 / \sqrt{3}}\Big) ,\quad \varphi=\Big(\frac{y_1}{1 +y_2 / \sqrt{3}}\Big).$
 \end{center}
 \noindent
The superpotential transforms non-trivial between these two basis. Specifically we consider:
 \begin{center}
$ W(T,\varphi) \rightarrow \bar{W}(y_1,y_2)= (1+y_2 /\sqrt{3})^3W.$
\end{center}

In order to find Starobinsky-like effective scalar potential we consider two forms of superpotentials which are written first in the more revealing basis $(y_1,y_2)$ and then in basis $(T,\varphi)$:
\begin{equation}
 W_{WZ}=\Big(\frac{\hat \mu}{2} \Big(y_1^2 +\frac{y_1^2 y_2}{\sqrt{3}}\Big) -\lambda \frac{y_1^3}{3}\Big ) \iff	 W'_{WZ}= \frac{\hat \mu}{2} \varphi^2  - \frac{\lambda}{3} \varphi^3 
 \label{w_1staro}
\end{equation}

\begin{equation}
 W_{C}=m\Big(-y_1y_2 +\frac{y_2y_1^2}{l\sqrt{3}}\Big)  \iff W'_{C}=\sqrt{3}m \varphi\left(T- \frac{1}{2}\right).
  \label{w_2staro}
\end{equation}
In the $(T,\varphi)$ basis one can recognize that Eq.~(\ref{w_1staro}) is the Wess-Zumino superpotential and the Eq.~(\ref{w_2staro}) is the Cecotti one~\cite{Ellis:2013xoa,Ellis:2018zya}. In order to derive Starobinsky-like effective scalar potential from the above superpotials, we consider that the inflaton field is the $y_1$ and the modulo the $y_2$.
If we chance basis the role of inflaton for the $W'_{WZ}$ is the $\varphi$ and the corresponding for $W'_{C}$ is the $T$. Moreover, we should consider the following choices $\lambda / \mu =1/3$ and $\hat{\mu}=\mu/\sqrt{3}$ for $W_{WZ}$ and $l=1$ for $W_{C}$.
%
The superpotentials in $(y_1,y_2)$ basis keep the transformations laws:
\begin{equation}
y_1 \rightarrow \alpha y_1 + \beta y_2, \quad y_2 \rightarrow - \beta^* y_1 +\alpha ^* y_2.
\end{equation}
where $\alpha,\beta $ are complex numbers and $|\alpha|^2+|\beta|^2=1$ \cite{Ellis:2018zya,Ellis:2013nxa}.

\begin{figure}
\begin{center}
\includegraphics[width=35pc]{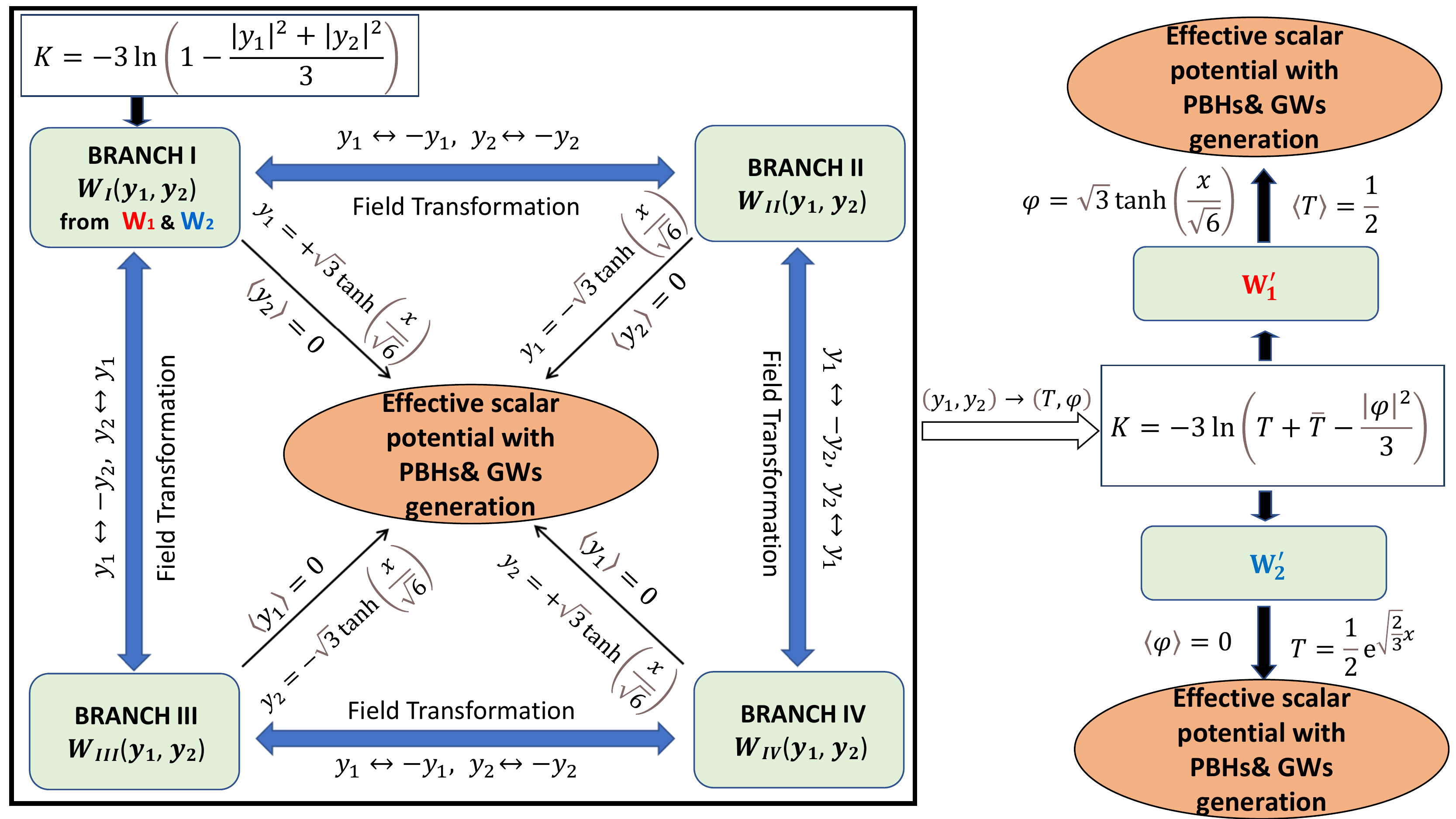}
\end{center}
\caption{\label{figure1}Figure adapted from Ref.~\cite{Ellis:2018zya}.}
\end{figure}

However, Starobinsky-like effective scalar potential cannot explain the generation of PBHs and GWs. As the detection of GWs by LIGO/VIRGO and the future space based experiments for GWs are on the topic of the cosmology we consider the following two superpotentials, which come from a modification of  Eqs.~(\ref{w_1staro}) and (\ref{w_2staro})\cite{Stamou:2021qdk}:

\begin{equation}
 W_1=\Big(\frac{\hat \mu}{2} \Big(y_1^2 +\frac{y_1^2 y_2}{\sqrt{3}}\Big) -\lambda \frac{y_1^3}{3}\Big ) (1+e^{-b_1y_1^2}(c_1 y_1^2+c_2y_1^4))
 \label{eqw1}
\end{equation}

\begin{equation}
W_2=m\Big(-y_1y_2 +\frac{y_2y_1^2}{l\sqrt{3}}\Big)(1+c_3e^{-b_2y_1^2}y_1^2) .
 \label{eqw2}
\end{equation}
These superpotentials are written in the $(y_1,y_2)$ basis. In the ($T,\varphi$) they take the form:
\begin{equation}
W_1'=\left( \frac{\hat{\mu}}{2}\varphi^2 -\frac{\lambda}{3} \varphi^3\right)\left(  1+ e^{-b_1 \left( \frac{2 \varphi}{1+ 2 T} \right)^2} \left[ c_1 \left( \frac{2 \varphi}{1+ 2 T} \right)^2+ c_2 \left( \frac{2 \varphi}{1+ 2 T} \right)^4\right] \right)
\label{wcase3}
\end{equation}

\begin{equation}
W_2'= \frac{\sqrt{3}}{2} m \varphi \Big( \frac{1}{2} -T\Big) \left( -1-2T -\frac{1-2T}{l} \right) \left[ 1+ 3c_3 e^{-3b_2 \left( \frac{1-2T}{1+2T} \right)^2}\left( \frac{1-2T}{1+2T} \right)^2\right].
\label{wcase4}
\end{equation}

In figure~\ref{figure1} we notice that starting by the K\"ahler potential in $(y_1,y_2)$ basis, the effective-scalar potential can keep its form under the field transformation colored with blue arrows. In order to find the effective scalar potential we need to set which is the inflaton field and which is the modulo and we need fix the non canonical kinetic term  for the inflaton by a further redefinition of the field, which is depicted in this figure with black inner arrows.  We can obtain the same effective effective scalar potential throughout the cycle in the left side of this figure by interplay the role of inflaton and modulo field according to this figure. Moreover, we can obtain the same effective scalar potential if we move to the other basis $(T,\varphi)$ and consider the other form of K\"ahler potential and the corresponding forms of superpotential from Eqs.~(\ref{wcase3}) and (\ref{wcase4}). This study and the importance of the unification between the models have previously analyzed for the case of Starobinsky scalar potential by considering Eqs.~(\ref{w_1staro}) and (\ref{w_2staro})\cite{Ellis:2013xoa,Ellis:2018zya,Ellis:2013nxa}. However these two modified superpotentials~(\ref{eqw1}) and (\ref{eqw2}) (or their equivalent forms (\ref{wcase3}) and (\ref{wcase4})) can explain the generation of PBHs and GWs due to the inflection point, which is presented in their effective scalar potential.

\section{Evaluating Scalar Power Spectrum}
In this section we present our analysis in order to derive numerically the scalar power spectrum. The equation of motion for the inflaton field is given in efold times by the expression:
\begin{equation}
\label{eq21}
\chi'' +3 \chi'- \frac{1}{2} \chi'^3 +\left(3- \frac{1}{2} \chi'^2\right) \frac{d\ln V(\chi)}{d \chi}=0
\end{equation}  
where with primes we denote the derivative in efold time  and as the equations, which follow, are valid for canonical kinetic term in the Lagrangian. In figure~\ref{figure1}, one can notice how the canonical $\chi$ field is connected to the $(y_1,y_2)$ fields and the $(T,\varphi)$ each time.

If we consider that the perturbation of the field is given as $\chi+ \delta \chi$ we have for the field's perturbation the following expression:
\begin{equation}
 \label{eq1.2} 
\delta\chi''=- \left(  3-\frac{1}{2}  \chi'^2 \right) \delta \chi'-\frac{1}{H^2}\frac{d^2V}{d\chi^2} \delta \chi -\frac{k^2}{a^2 H^2} \delta \chi +4 \Psi' \chi'-\frac{2 \Psi}{H^2}\frac{dV}{d\chi}
\end{equation}
and  for the Bardeen potential $\Psi$ we have the equation:
\begin{equation}
 \label{eq1.3}
\Psi''=-\left(  7-\frac{1}{2} \chi'^2 \right) \Psi'- \left( 2\frac{V}{H^2} +\frac{k^2}{a^2 H^2}\right)\Psi -\frac{1}{H^2}\frac{dV}{d\chi} \delta \chi
\end{equation} 
where $k$ is the comoving wavenumber and $H$ is the Hubble parameter. 

After integrating  Eqs.(\ref{eq21}), (\ref{eq1.2}) and (\ref{eq1.3}) for each comoving wavenumber $k$, we can evaluate the scalar power spectrum from the following expression:
\begin{equation}
\label{eq1.5}
P_R=\frac{k^3}{2 \pi^2} \left|R_k \right|^2,
\end{equation}
where $R_k$ is the comoving curvature perturbation:
\begin{equation}
R_k=\Psi +\frac{\delta \phi }{\phi'}.
\label{eq1.6}
\end{equation}
An analysis of the numerical treatment can be found  in Ref.\cite{Ringeval:2007am,Nanopoulos:2020nnh}.

\begin{figure}
\begin{center}
\includegraphics[width=30pc]{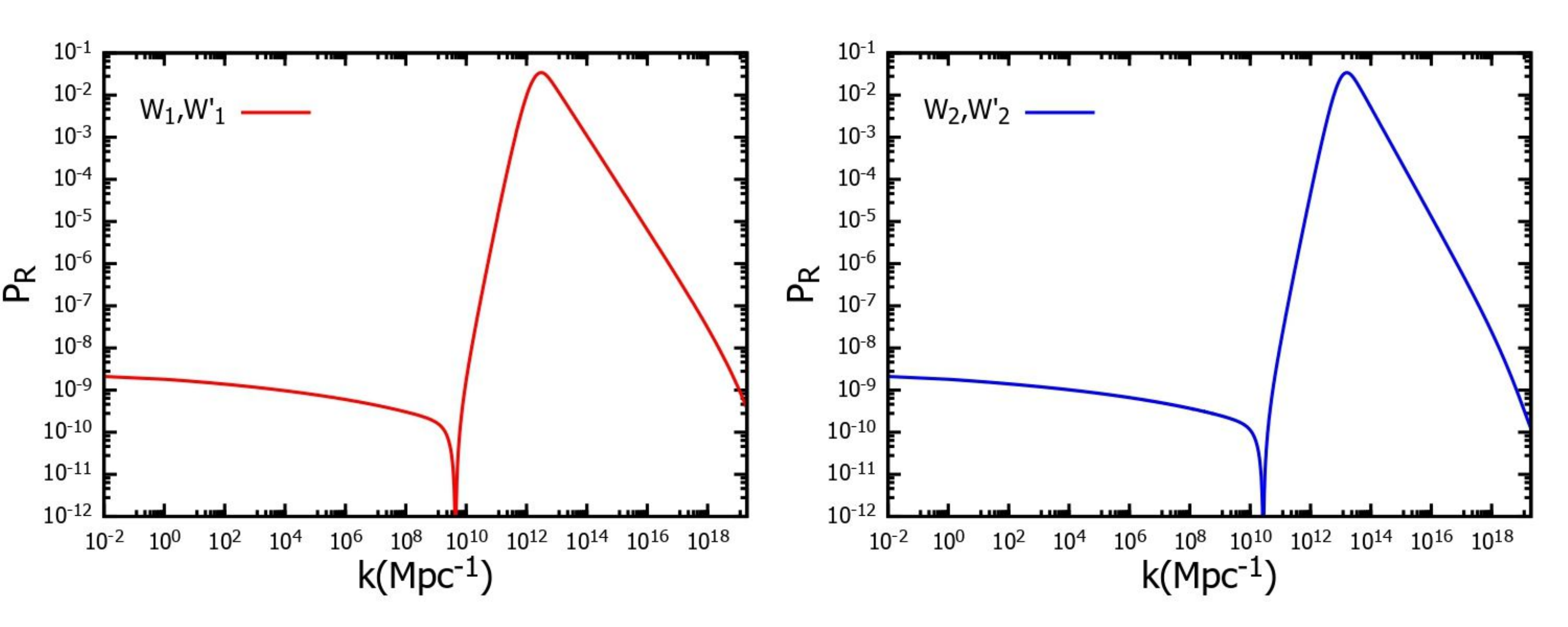}
\end{center}
\caption{\label{figure2} Scalar power spectrum for the superpotenrial $W_1$ and $W_2$ (or $W'_1$ and $W'_2$). Parameters: $c_1=1.7$, $c_2=-0.65$, $c_3=14$, $b_1=3.701266$, $b_2=8.995722$, $l=1.0002$, $\lambda / \mu =1/3$.}
\end{figure}
%

In figure~\ref{figure2} we show the scalar power spectrum for the superpotentials $W_1$ and $W_2$, given from Eqs.~(\ref{eqw1}) and (\ref{eqw2}), with red and blue respectively.  We can notice that we have an significant enhancement of power at around $10^{13}$ $Mpc^{-1}$. This enhancement can be interpreted as the production of PBH and GWs in radiation dominated epoch with the origin to the previous epoch of inflation. Moreover it can be interpreted as a significant fraction of the DM in the Universe. The results for their equivalent forms $W_1'$ and $W_2'$, given from Eqs.~(\ref{wcase3}) and (\ref{wcase4}) are the same. For initial condtions of the field we choose $\chi_{ic}=4.95$ for $W_1$ and $\chi_{ic}=4.87$ for $W_2$.
\section{Production of PBHs}

The present abundance of PBHs is given by the integral:
\begin{equation}
f_{PBH} = \int d\ln M \frac{\Omega_ {PBH}}{\Omega_ {DM}}
\end{equation}
where
\begin{equation}
\label{44}
\frac{\Omega_ {PBH}}{\Omega_ {DM}}= \frac{\beta(M(k))}{8 \times 10^{-16}} \left(\frac{\gamma}{0.2}\right)^{3/2} \left(\frac{g(T_f)}{106.75}\right)^{-1/4}\left(\frac{M(k)}{10^{-18} \textsl{g}}\right)^{-1/2}.
\end{equation}
The mass is given as a function of k mode:
\begin{equation}
\label{43}
M_{PBH}=10^{18}  \left(\frac{\gamma}{0.2}\right)  \left(\frac{g}{106.75}\right)^{-1/6} \left(\frac{k}{7 \times 10^{13} Mpc^{-1}  }\right)^{-2} \textsl{g}
\end{equation}
where $g$ is the effective number of degree of freedom and $\gamma$ is the correction factor which depends on the gravitational collapse. The mass fraction of the Universe collapse in PBH of $M_{PBH}$, denoted as $\beta$, is calculated with two methods, the Press-Schecter approach (PS)  and the peak theory (PT).

In the PS approach the mass fraction $\beta_{PS}$ is given by the probability that the overdensity $\delta$ is above a certain threshold of collapse, denoted as $\delta_c$: 
\begin{equation}
\label{42}
\beta_{PS}(M_{PBH})= \frac{1}{\sqrt{2 \pi \sigma ^2 (M)}} \int^{\infty}_{\delta_c} d\delta ~ e^{\frac{-\delta ^2}{2 \sigma^2(M)}} =\frac{\Gamma\left(\frac{1}{2}, \frac{\delta_c^2}{2 \sigma^2}\right)}{2\sqrt{\pi}} .
\end{equation}
The variance of the curvature perturbation $\sigma$ is related to the power spectrum:
\begin{equation}
\label{40}
\sigma^2 (M_{PBH}(k))= \frac{4(1+\omega)^2}{(5+3\omega)^2}  \int \frac{dk' }{k'} \left(\frac{k'}{k}\right)^4 P_R(k') \tilde W^2\left(\frac{k'}{k}\right)
\end{equation}
where $\tilde W$ is a window function. Throughout this work we consider the Gaussian distribution for this function. With $\omega$ we denote the equation of state and as we work in radiation dominated epoch we consider that $\omega =1/3$.

In the PT, the number density of peaks above a threshold is given by the expression:
\begin{equation}
n_{peaks}= \frac{1}{(2 \pi)^2} \left(  \frac{\left\langle {{k^2}} \right\rangle}{3}\right)^{3/2}\left( \left(\frac{\delta_c}{\sigma}\right)^2-1\right)exp\left(  - \frac{\delta_c^2}{2 \sigma^2}\right)
\end{equation}
where $\left\langle {{k^2}} \right\rangle$ reads as:
\begin{equation}
\left\langle {{k^2}} \right\rangle= \frac{1}{\sigma^2} \int^{\infty}_{0} \frac{dk'}{k'} {k'}^2\tilde W \left(\frac{k'}{k}\right)^2 P_{\Delta}(k').
\end{equation}
The density power spectrum is defined as:
\begin{equation}
P_{\Delta}=\frac{4(1+\omega)^2}{(5+3\omega)^2} \left(\frac{k'}{k}\right)^4P_R(k').
\end{equation}
The mass fraction for a flat Universe  assuming a  Gaussian  distribution is given by the expression:
\begin{equation}
\beta_{PT}=n_{peaks}(2\pi)^{3/2}(1/k)^3.
\end{equation}

\begin{figure}
\begin{center}
\includegraphics[width=30pc]{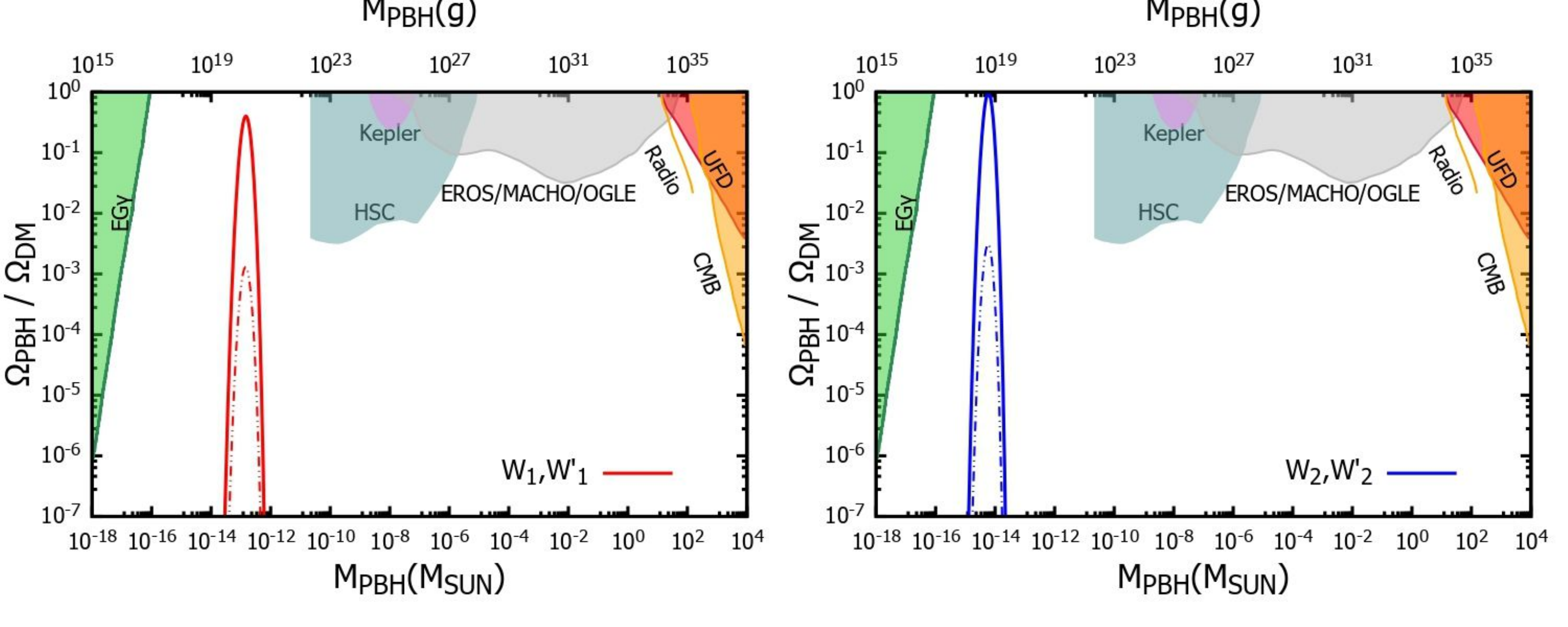}
\end{center}
\caption{\label{figure3} Fractional abundance of PBH for $W_1$ and $W_2$ (or $W'_1$ and $W'_2$).}
\end{figure}
In figure~\ref{figure3} we present the abundances derived by the superpotentials $W_1$ (red) and $W_2$ (blue), or  their equivalent forms $W_1'$ and $W_2'$. Solid lines correspond to the PT  and dashed line to PS. We  notice that there is a difference between these two methods.  However this difference can be compensated by the fact that the result are sensitive to the value of $\delta_c$. In this figure we consider that $\delta_c=0.45$. The choice of parameters are the same with the figure \ref{figure2}. In this figure we also depict for comparison reason the observable bounds that come from extragalactic gamma ray from PBH evaporation ($EG\gamma$),  microlensing for Subaru (HSC), Eros/Macho/Ogle and Kepler dynamical heating of ultra faint dwarf (UFD), CMB measurements  and radio observation.

In table~\ref{tabu1} we show the fractional abundances for the two proposed models $W_1$ and $W_2$. We have notice that the values for fractional abundance using PT are bigger than those of PS, as we previously observe by the figure~\ref{figure3}. In this table we show the value of the peak of the power spectrum and the value for the observable constraints of $n_s$ and $r$. It is important to remark here that our proposed models are in complete consistence with the observable constraints released by Planck collaboration 2018\cite{Akrami:2018odb}. The $n_s$ and $r$ are calculated at leading order in the slow-roll expansion from the following expressions:
\noindent
\begin{equation}
\label{21a}
n_s \simeq 1+2 \eta_V -6\varepsilon_V, \quad r \simeq 16 \varepsilon_V,
\end{equation} 
where the slow-roll $\varepsilon_V$ and $\eta_V$ parameters are given by the following expression:
\begin{equation}
\varepsilon_V= \frac{1}{2}\left(\frac{V'(\chi)}{V(\chi)} \right)^2 , \quad \eta_V=\frac{V''(\chi)}{V(\chi)}.
\end{equation}

\begin{center}
\begin{table}[h!]
\caption{The value for the $n_s$, $r$}
\centering
\begin{tabular}{@{}*{7}{l}}
\br
 & $n_s$& $r$ & $P_R^{peak}$& $f_{PBH}^{PS}$ &$f_{PBH}^{PT}$\\
\mr
1 & $0.9634$& $0.0099$ & $0.036$& $ 0.00239$& $0.752$\\
2 &$0.9605$& $0.0127$& $0.034$ &$0.00201$ &$ 0.651$\\
\label{tabu1}
\end{tabular}
\end{table}
\end{center}
\section{Production of GWs}
The energy density of the GWs in terms of scalar power spectrum is given by thr integral:
\begin{equation}
{\Omega_{GW}(k)}=\frac{c_g\Omega_r}{36} \int^{\frac{1}{\sqrt{3}}}_{0}\mathrm{d} d
\int ^{\infty}_{\frac{1}{\sqrt{3}}}\mathrm{d} s \left[  \frac{(s^2-1/3)(d^2-1/3)}{s^2+d^2}\right]^2\times
P_{R}(kx)P_{R}(ky)(I_c^2+I_s^2)
\label{eq4.1}
\end{equation}
where the radiation density is $\Omega_r \approx 5.4 \times 10^{-5}$.
The variables $x$ and $y$ are defined as:
\begin{equation}
x= \frac{\sqrt{3}}{2}(s+d), \quad  y=\frac{\sqrt{3}}{2}(s-d).
\label{eq4.2}
\end{equation}
Finally, the functions $I_c$ and $I_s$ are given by the equations:
\begin{gather}
I_c=-36 \pi \frac{(s^2+d^2-2)^2}{(s^2-d^2)^3}\theta(s-1)\\
I_s=-36 \frac{(s^2+d^2-2)^2}{(s^2-d^2)^2}\left [ \frac{(s^2+d^2-2)}{(s^2-d^2)} log\left| \frac{d^2-1}{s^2-1} \right| +2\right].
\label{eq4.3}
\end{gather}


\begin{figure}
\begin{center}
\includegraphics[width=30pc]{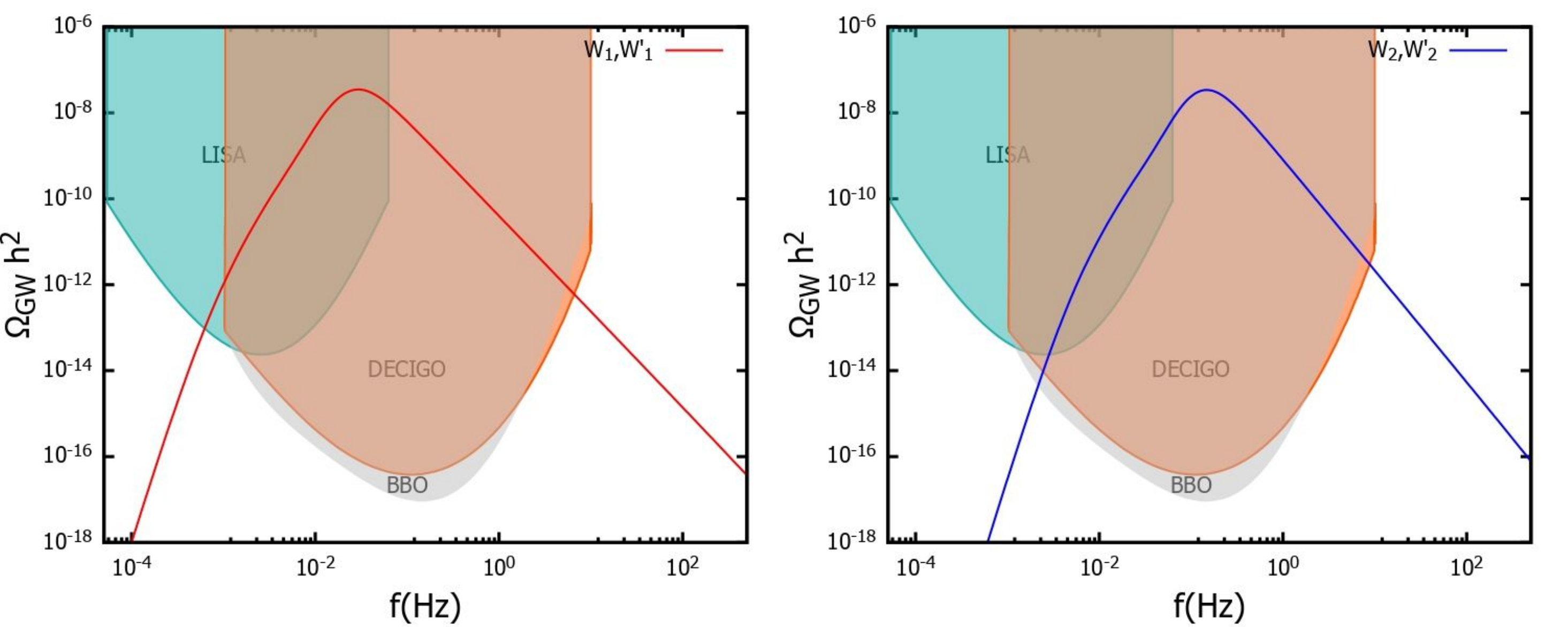}
\end{center}
\caption{\label{figure4} Energy density of GWs for $W_1$ and $W_2$ (or $W'_1$ and $W'_2$).}
\end{figure}

In figure~\ref{figure4} we present the energy density for GWs for the superpotentials $W_1$ (red) and $W_2$ (blue), or for their equivalent forms $W_1'$ and $W_2'$. The choice of parameters are the same with  figure~\ref{figure2}, as in the study of PBHs. We plot them in respect to the frequency in order to have a comparison with future space-based experiments, such as LISA, DECIGO and BBO. We can notice that both proposed models $W_1$ and $W_2$ give us significant results for the amount of GWs.
\section{Fine-tuning}
\begin{figure}
\begin{center}
\includegraphics[width=30pc]{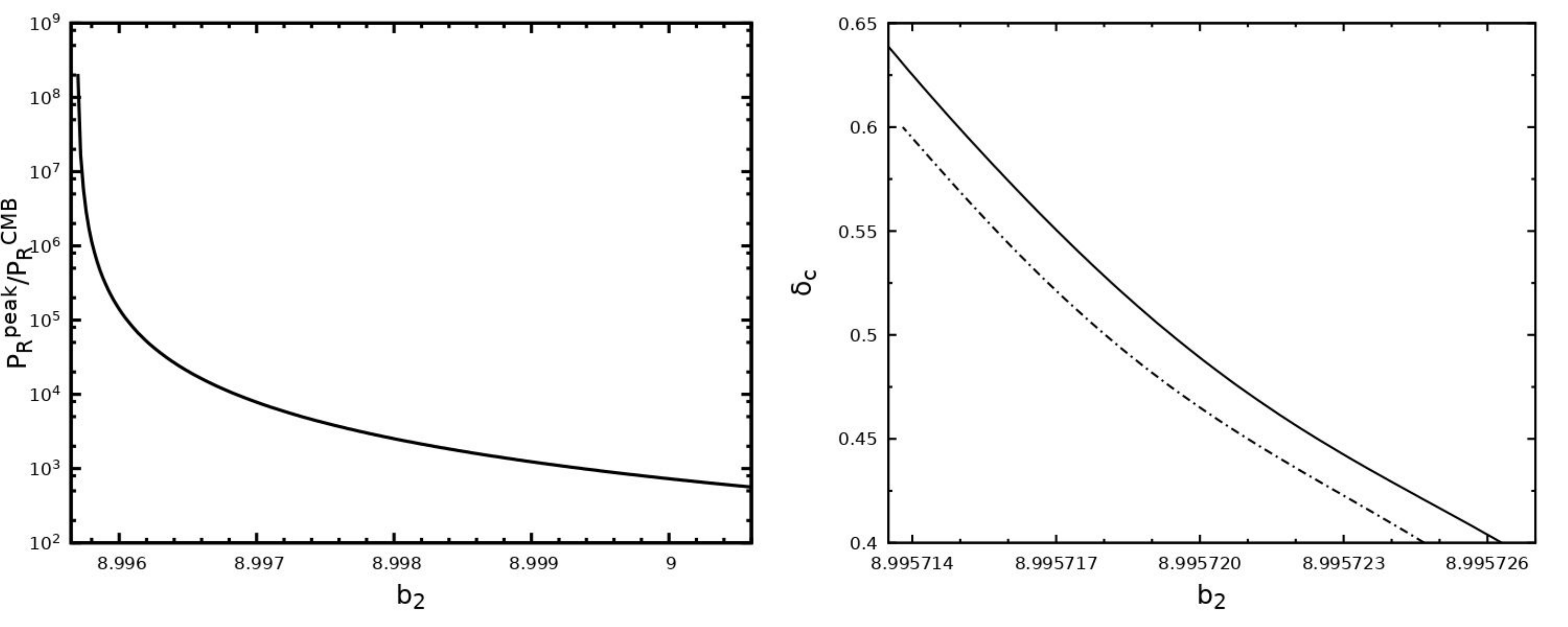}
\end{center}
\caption{\label{figure5} How $b_2$ releated to the enchantment of power spectrum and to the $\delta_c$.}
\end{figure}

Finally, the fine-tuning is regarded as a drawback in models with an inflection point. In our analysis the parameters that need fine tuning are  $b_{1,2}$. In  figure~\ref{figure5} we depict how the value of $b_2$, for instance, which comes  from $W_2$,   depends on the enhancement of the power spectrum. In the case of the production of PBHs  a peak at least $10^{-2}$  is required. Specifically, the value of the parameters $b_{1,2}$ which are related of the value of the peak need to take such values in order to use the acceptable threshold $\delta c$. As we see in the right panel of  figure~\ref{figure5}, if we want to use the value of $\delta c$ at a range between $ [0.4 - 0.6] $,  we should fix the value of  $b_{2}$ according to this plot. This plot is made assuming that the fractional abundance of PBHs is $f_{PBH}=0.1$. We numerically evaluate the value of  $b_2$ by repeating many times the calculation for PBHs production. Specifically, we define a value for $\delta c$ and we stop the running when the $f_{PBH}$ reaches the value $0.1$. The dashed line corresponds to PS approach and the solid to the PT. We can observe again a difference between these two approaches.

As a next step in order to have a measurement for the required fine tuning we evaluate the quantity of $\Delta $, which  is the maximum value of the following logarithmic derivative:
\begin{equation}
\Delta=\mathrm{Max} \left| \frac{\partial ln(P_R^{PEAK}) }{\partial ln(b_i) }   \right|.
\end{equation}
If the max value of this logarithmic derivative is big, we need a lot of fine tuning. In case of studying PBHs the value of $\Delta$ is up to $10^6$. Hence a lot of fine tuning is required. This is due to the value of $\delta_c$, which should be $\delta_c=0.4-0.6$.

However in the study of GWs the required fine-tuning can be decreased because there is not the restriction of $\delta c$ and because the peak of scalar power spectrum should not be at least $10^{-2}$. Therefore, if we demand a lower peak, this quantity $\Delta$ can be decreased by two or three order of magnitude.

\section{Conclusions}
In this work we provide a class of superpotential in order to generate PBHs and GWs, which is derived, by no-scale supergravity. We study superpotentials using the non-compact SU(2,1)/SU(2)$\times$ U(1) symmetry.  By a proper modification in the superpotentials, which in unmodified case lead to Starobinsky-like effective scalar potential, we derive potentials which are capable not only to generate PBHs and GWs, but also to obtain the same transformation laws in $(y_1,y_2)$ and  keep the form of effective scalar potential in both basis $(T,\varphi)$ and $(y_1,y_2)$. The machanism in order to achieve enhancement in scalar power spectrum and  explain the production of PBHs and GWs is an inflection point in the effective scalar potential.

As a next step,  we evaluate the production of PBHs by using two approaches the PS approach and the PT.  Furthermore, we evaluate the abundances GWs by using the scalar power spectra. Our proposed models give us significant results in order to expain the production of PBHs and the DM in the Universe as well the generation of GWs. For both the production of PBHs and GWs we assume that they are formed in radiation dominated epoch.

Finally, we discuss the issue of fine-tuning of the parameters of the effective scalar potential. We conclude that the apparent drawback of models with an inflection point is that it is required a lot of fine-tuning. Therefore, other theoretical approaches, such as multi-field inflation, should be taken into consideration in the study of PBHs and GWs.


\section*{Acknowledgments}
  \label{acknow}
The research work was supported by the Hellenic Foundation for Research
and Innovation (H.F.R.I.) under the First Call for H.F.R.I. Research Projects to support Faculty members and
Researchers and the procurement of high-cost research equipment grant  (Project Number: 824). I.D.S. would like to thank V.C. Spanos for useful discussions.


\section*{References}
\begin{thebibliography}{9}


\bibitem{Abbott:2016blz}
B.~P.~Abbott \textit{et al.} [LIGO Scientific and Virgo],
Phys. Rev. Lett. \textbf{116} (2016) no.6, 061102
doi:10.1103/PhysRevLett.116.061102
[arXiv:1602.03837 [gr-qc]].


\bibitem{Abbott:2016nmj}
B.~P.~Abbott \textit{et al.} [LIGO Scientific and Virgo],
Phys. Rev. Lett. \textbf{116} (2016) no.24, 241103
doi:10.1103/PhysRevLett.116.241103
[arXiv:1606.04855 [gr-qc]].

\bibitem{Abbott:2017vtc}
B.~P.~Abbott \textit{et al.} [LIGO Scientific and VIRGO],
Phys. Rev. Lett. \textbf{118} (2017) no.22, 221101
[erratum: Phys. Rev. Lett. \textbf{121} (2018) no.12, 129901]
doi:10.1103/PhysRevLett.118.221101
[arXiv:1706.01812 [gr-qc]].

%

\bibitem{Akrami:2018odb}
Y.~Akrami \textit{et al.} [Planck],
Astron. Astrophys. \textbf{641} (2020), A10
doi:10.1051/0004-6361/201833887
[arXiv:1807.06211 [astro-ph.CO]].


\bibitem{Ellis:2013xoa}
J.~Ellis, D.~V.~Nanopoulos and K.~A.~Olive,
Phys. Rev. Lett. \textbf{111} (2013), 111301
[erratum: Phys. Rev. Lett. \textbf{111} (2013) no.12, 129902]
doi:10.1103/PhysRevLett.111.111301
[arXiv:1305.1247 [hep-th]].


\bibitem{Ellis:2018zya}
J.~Ellis, D.~V.~Nanopoulos, K.~A.~Olive and S.~Verner,
JHEP \textbf{03} (2019), 099
doi:10.1007/JHEP03(2019)099
[arXiv:1812.02192 [hep-th]].

\bibitem{Ellis:2013nxa}
J.~Ellis, D.~V.~Nanopoulos and K.~A.~Olive,
JCAP \textbf{10} (2013), 009
doi:10.1088/1475-7516/2013/10/009
[arXiv:1307.3537 [hep-th]].

\bibitem{Cremmer:1983bf}
E.~Cremmer, S.~Ferrara, C.~Kounnas and D.~V.~Nanopoulos,
Phys. Lett. B \textbf{133} (1983), 61
doi:10.1016/0370-2693(83)90106-5

\bibitem{Ellis:1983ei}
J.~R.~Ellis, C.~Kounnas and D.~V.~Nanopoulos,
Nucl. Phys. B \textbf{241} (1984), 406-428
doi:10.1016/0550-3213(84)90054-3


\bibitem{Stamou:2021qdk}
I.~D.~Stamou,
Phys. Rev. D \textbf{103} (2021) no.8, 083512
doi:10.1103/PhysRevD.103.083512
[arXiv:2104.08654 [hep-ph]].

\bibitem{Ringeval:2007am}
C.~Ringeval,
Lect. Notes Phys. \textbf{738} (2008), 243-273
doi:10.1007/978-3-540-74353-8\_7
[arXiv:astro-ph/0703486 [astro-ph]].

\bibitem{Nanopoulos:2020nnh}
D.~V.~Nanopoulos, V.~C.~Spanos and I.~D.~Stamou,
Phys. Rev. D \textbf{102} (2020) no.8, 083536
doi:10.1103/PhysRevD.102.083536
[arXiv:2008.01457 [astro-ph.CO]].

\end{thebibliography}
\end{document}